\pdfoutput=1

\documentclass[twocolumn,aps,prb,superscriptaddress]{revtex4}

\usepackage{amssymb,amsfonts,amsmath}
\usepackage{color,graphicx}
\usepackage[normalem]{ulem}

\newcommand{\R}{\mathbb R}

\newcommand{\m}{\mathbf{m}}
\newcommand{\n}{\mathbf{n}}
\newcommand{\mpa}{m_\parallel}
\newcommand{\mpaO}{m_{0,\parallel}}
\newcommand{\mpaI}{m_{1,\parallel}}

\newcommand{\mpe}{\mathbf{m}_\perp}
\newcommand{\mpeb}{\overline{\mathbf{m}}_\perp}
\newcommand{\mpeO}{\mathbf{m}_{0,\perp}}
\newcommand{\mpeI}{\mathbf{m}_{1,\perp}}

\begin{document}

\title{Theory of Dzyaloshinskii domain wall tilt in ferromagnetic
  nanostrips}

\author{Cyrill B. Muratov}

\affiliation{Department of Mathematical Sciences, New Jersey Institute
  of Technology, Newark, NJ 07102, USA}

\author{Valeriy V. Slastikov}

\affiliation{School of Mathematics, University of Bristol, Bristol BS8
  1TW, United Kingdom}

\author{Alexander G. Kolesnikov}

\affiliation{School of Natural Sciences, Far Eastern Federal
  University, Vladivostok 690950, Russia}

\author{Oleg A.  Tretiakov}
\email{olegt@imr.tohoku.ac.jp}
\affiliation{Institute for Materials Research, Tohoku University,
  Sendai 980-8577, Japan}
\affiliation{School of Natural Sciences, Far Eastern Federal
  University, Vladivostok 690950, Russia}

\date{June 21, 2017}

\begin{abstract}
  We present an analytical theory of domain wall tilt due to a
  transverse in-plane magnetic field in a ferromagnetic nanostrip with
  out-of-plane anisotropy and Dzyaloshinskii-Moriya interaction
  (DMI). The theory treats the domain walls as one-dimensional objects
  with orientation-dependent energy, which interact with the sample
  edges. We show that under an applied field the domain wall remains
  straight, but tilts at an angle to the direction of the magnetic
  field that is proportional to the field strength for moderate fields
  and sufficiently strong DMI. Furthermore, we obtain a nonlinear
  dependence of the tilt angle on the applied field at weaker DMI. Our
  analytical results are corroborated by micromagnetic simulations.
\end{abstract}

\maketitle

\section{Introduction}
\label{sec:introduction}

Domain wall (DW) statics and dynamics in thin-film ferromagnetic
systems have been a subject of intense
experimental\cite{Atkinson03,Yamaguchi04, Parkin:racetrack08,
  Hoffmann2015_review} and theoretical \cite{Tatara04, Thiaville05,
  Tretiakov08,Fert2013, Shibata2011_review} studies over the last
decades due to their direct relevance to spintronic
memory\cite{Parkin08} and logic devices. \cite{Allwood02} Recently, it
has been realized that ferromagnets with Dzyaloshinskii-Moriya
interaction\cite{Dzyaloshinskii58,Moriya60} (DMI) may offer more
benefits in this direction, \cite{Tretiakov_DMI, Thiaville2012,
  Boulle13} which led to an enormous experimental progress for these
systems. \cite{Emori2014,Franken2014,
  Marrows_SAF2015,Boulle2016,Jiawei2016}

Due to their better technological suitability as smaller and more
robust carriers of information in spintronic nanodevices, the DWs in
ultrathin ferromagnetic films with out-of-plane anisotropy and
interfacial DMI have now become the primary objects of experimental
interest. \cite{Emori2014,Franken2014,Marrows_SAF2015,Boulle2016,Jiawei2016}
Moreover, it was discovered that the DWs move much more efficiently in
these systems due to spin-orbit torques.\cite{Garello2013, Fert2013,
  Ado2017} It was also demonstrated that in these systems the DW
equilibrium structure changes from Bloch to N\'eel type in the
presence of strong DMI. \cite{chen13} Following a theoretical
study,\cite{Thiaville2012} we refer below to this new type of magnetic
DWs as {\em Dzyaloshinskii} domain walls.

Boulle \textit{et al.}\cite{Boulle13} were the first to discover
numerically that these DWs develop a tilt under in-plane magnetic
fields and applied currents. This DW tilt was shown to depend on the
DMI and field strengths. It was followed by several more attempts to
investigate this phenomena
theoretically\cite{Martinez2014,Vandermeulen2016} and multiple
experimental studies.\cite{Emori2014,Franken2014,Jiawei2016} However,
up to now there is still lack of a unifying theory of the DW tilt and
its dependence on the in-plane magnetic field.

\begin{figure}
  \centering
  \includegraphics[width=3.25in]{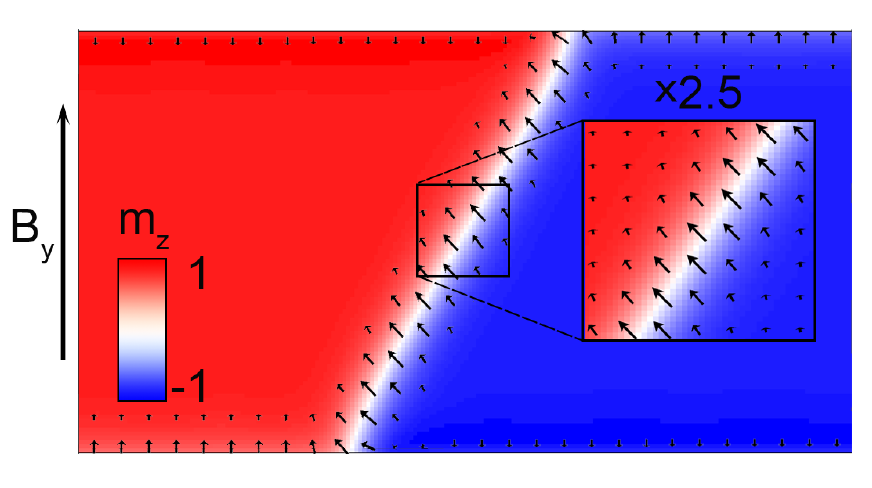}
  \caption{An example of a tilted domain wall in a transverse in-plane
    magnetic field computed using micromagnetic simulations.  The
    material parameters are: $A = 10^{-11}$J/m, $K = 1.25 \times 10^6$
    J/m$^3$, $M_s = 1.09 \times 10^6$ A/m, $D = 2.4$ mJ/m$^2$ and
    $B_y = \mu_0 H = 80$ mT (see Sec.~\ref{sec:model} for precise
    definitions).  }
  \label{fig:tilt2dnum}
\end{figure}

In this paper, we explain these important findings on more solid
theoretical grounds, using variational analysis of the magnetic energy
functional. We study a DW in a thin nanostrip with perpendicular
magnetic anisotropy (PMA) and interfacial DMI in the presence of a
magnetic field applied in the plane of the strip and perpendicular to
its axis, see Fig.~\ref{fig:tilt2dnum} for an example. For this system
we analytically develop a proper reduced geometric variational model
built on exact one-dimensional (1D) DW solutions to describe the
equilibrium tilted two-dimensional (2D) DW configurations. These exact
1D magnetization profiles are in general neither N\'eel nor Bloch
type, but can be easily computed numerically for all relevant values
of the parameters. We also derive explicit analytical expressions by
expanding the DW energy in the applied field or DMI strengths.

In the reduced 2D variational problem, we treat the DW as a curve
whose shape is determined by minimizing an appropriate geometric
energy functional.  As a result we show that the DW in equilibrium
remains straight despite the fact that the wall energy is a function
of its local orientation. In particular, for small fields the tilt
angle is found to be proportional to the transverse magnetic field
strength.  One of the features of our 2D analysis is the necessity to
include the edge DWs found earlier in the context of
skyrmions.\cite{Rohart2013} These edge DWs can be seen in
Fig.~\ref{fig:tilt2dnum} along the upper and lower strip edges. We
show that the contribution of the edge DWs is also essential for
determining the proper tilt angle.  This is because the total DW
energy contains contributions from both the internal and the edge DWs,
and it is the competition among them that determines the tilt
angle. We find that the effect of the edge DWs becomes weaker when the
DMI strength is reduced, whereas the internal DW energy has a
nontrivial dependence on the DMI, magnetic field, and wall orientation
that have to be properly accounted to determine the equilibrium tilt
angle.

The main advantage of our reduced geometric variational model for
tilted DWs is its considerable simplicity compared to the full
micromagnetic description. Specifically, it allows for a detailed
analytical treatment, which highlights the key physical features of
tilted DWs mediated by interfacial DMI in PMA nanostructures. In
particular, it yields explicit closed-form expressions for the
dependence of the equilibrium tilt angle for a wide range of the
material parameters and applied fields. The obtained analytical
predictions are found to be in excellent agreement with the results of
micromagnetic simulations, indicating that the reduced model captures
all the essential physical aspects of the considered system.

The paper is organized as follows. In Sec.~\ref{sec:model} we
introduce the full micromagnetic model and its 2D reduction
appropriate for infinite ultrathin ferromagnetic nanostrips. In
Sec.~\ref{sec:monodomain-states} the theory of edge domain walls is
presented, and in Sec.~\ref{sec:interior-wall} a detailed analysis of
1D interior wall profiles is carried out. Next, in
Sec.~\ref{sec:two-dimens-probl} we demonstrate how the theory of 1D
domain walls developed in the preceding sections is applied to a
Dzyaloshinskii DW in an infinite 2D nanostrip. In
Sec.~\ref{sec:comp-with-numer} we compare our analytical theory with
micromagnetic simulations and show a good agreement between them. Here
the additional effect of dipolar interactions is also discussed.
Finally, a summary and some concluding remarks are presented in
Sec.~\ref{sec:conclusions}.

\section{Model}
\label{sec:model}

We consider a thin ferromagnetic nanostrip exhibiting PMA and
interfacial DMI under the influence of an in-plane magnetic field. We
start with a three-dimensional micromagnetic energy
\cite{Hubert:book,Bogdanov89,Fert90,Bogdanov94} (in the SI units):
\begin{align}
  \label{Ephys}
  E(\mathbf M) 
  & =\int_\Omega \left(  {A \over M_s^2} 
    |\nabla \mathbf M|^2 +  {K \over M_s^2} |\mathbf M_\perp|^2 
    - \mu_0 
    \mathbf M \cdot \mathbf H \right) d^3 r \notag \\
  & + \mu_0 \int_{\R^3} \int_{\R^3}
    {\nabla \cdot \mathbf M(\mathbf r) \, \nabla \cdot \mathbf
    M(\mathbf r') \over 8 \pi | \mathbf r - \mathbf r'|} \, d^3 r
    \, d^3 r' \notag \\
  & + {D d \over M_s^2} \int_{\partial \Omega_0} \Big(
    \overline{M}_\parallel \nabla \cdot \overline{\mathbf 
    M}_\perp - \overline{\mathbf M}_\perp \cdot \nabla
    \overline{M}_\parallel \Big) d^2 r.    
\end{align}
Here $\mathbf M = \mathbf M(\mathbf r)$ is the magnetization vector at
point $\mathbf r = (x,y,z) \in \Omega \subset \R^3$, where
$\Omega = (-L/2, L/2) \times (-W/2, W/2) \times (0, d)$ is the
nanostrip of length $L$, width $W$ and thickness $d$, and
$\mathbf M_\perp$ and $M_\parallel$ are the in-plane and out-of-plane
components of $\mathbf M$, respectively. The terms in
Eq.~\eqref{Ephys} are, respectively: the exchange, uniaxial
perpendicular anisotropy, Zeeman, magnetostatic interactions and the
interfacial DMI terms, and $M_s = |\mathbf M|$, $A$, $K$, $\mathbf H$
and $D$ are the saturation magnetization, exchange stiffness,
anisotropy constant, applied magnetic field and the DMI strength. As
usual, $\mu_0$ is the permeability of vacuum. In the magnetostatic
energy term, the vector field $\mathbf M(\mathbf r)$ is extended by
zero outside $\Omega$, and $\nabla \cdot \mathbf M$ is understood
distributionally (i.e., it includes the contributions of boundary
charges). Since the considered DMI is due to interfacial effects, its
contribution to the energy is via a surface integral over the bottom
film surface $\partial \Omega_0$ corresponding to an interface between
the ferromagnet and a heavy metal, and
$\overline{\mathbf M} = (\overline{\mathbf M}_\perp, \overline
M_\parallel)$
is the value of $\mathbf M$ on $\partial \Omega_0$. However, using the
standard convention, we normalize the DMI strength parameter $D$ to a
unit volume of the ferromagnet.

We assume that the applied magnetic field is in the plane of the film
and is normal to the strip axis, i.e.,
$\mathbf H = H \hat{\mathbf y}$, where $\hat{\mathbf y}$ is the unit
vector in the direction of the $y$-axis. We also consider films which
are much thinner than the exchange length
$\ell_{ex} = \sqrt{2 A / (\mu_0 M_s^2)}$, so that the magnetization in
$\Omega$ is constant along the film thickness. Measuring lengths in
the units of $\ell_{ex}$ and setting
$\mathbf M(x, y, z) = M_s \m(x, y)$ with $|\m| = 1$ in $\Omega$, we
can rewrite the energy, to the leading order\cite{Gioia97} in
$d / \ell_{ex}$, in the units of $A d$ as
\begin{align}
  \label{Eh}
  E(\m) & \simeq \int_{-l/2}^{l/2} \int_{-w/2}^{w/2} \Big[ |\nabla \m|^2 +
          (Q - 1) |\mpe|^2 - 2 h \hat{\mathbf y} \cdot \mpe
          \notag \\
        & \quad + \kappa \left( \mpa \nabla \cdot \mpe -
          \mpe \cdot \nabla \mpa \right) \Big] \, dy \, dx.
\end{align}
Here we defined $\mpe \in \R^2$ and $\mpa \in \R$ to be the respective
in-plane and out-of-plane components of the unit magnetization vector
$\m$, introduced the dimensionless parameters
\begin{align}
  \label{Qkappa}
  Q = {2 K \over \mu_0 M_s^2}, \quad \kappa = D \sqrt{2 \over \mu_0
  M_s^2 A}, \quad h = {H \over M_s},
\end{align}
and defined the rescaled nanostrip dimensions $l = L / \ell_{ex}$ and
$w = W / \ell_{ex}$. In Eq.~\eqref{Qkappa}, $Q > 1$ is the material's
quality factor yielding PMA, $\kappa$ is the dimensionless DMI
strength, which without loss of generality, may be assumed positive,
and $h$ is the dimensionless applied field strength.

We are interested in the case of long nanostrips corresponding to
$l \gg w$. Note that when $l \to \infty$, the energy in Eq.~\eqref{Eh}
diverges even if $h = 0$ because of the presence of edge domain walls
giving $O(l)$ contribution to the energy.\cite{Rohart2013,Muratov2016}
Therefore, in order to pass to the limit $l \to \infty$ we need to
subtract from $E$ the contribution of the one-dimensional ground state
energy $e_0(h, w) = \min E_0(\m)$, where
\begin{align}
  \label{E0}
  E_0(\m) & = \int_{-w/2}^{w/2} \Big[ |\m'|^2 + (Q
            - 1) |\mpe|^2 - 2 h \hat{\mathbf y} \cdot \mpe \notag \\
          & \quad + \kappa \left( (\hat{\mathbf y} \cdot \mpe') \mpa -
            (\hat{\mathbf y} \cdot \mpe) \mpa'  \right) \Big] \, dy.
\end{align}
The precise functional form of $e_0(h, w)$ is the subject of
Sec.~\ref{sec:monodomain-states}.

Putting everything together, we now write the expression for the
energy that describes a Dzyaloshinskii domain wall running across the
nanostrip as
\begin{align}
  \label{E}
  E(\m) & = \int_{-\infty}^{\infty} \int_{-w/2}^{w/2} \Big[
          |\nabla \m|^2 + (Q - 1) |\mpe|^2  \notag \\
        & \quad - 2 h \hat{\mathbf y} 
          \cdot \mpe  - w^{-1} e_0(h,w) \notag \\
        & \quad + \kappa \left( \mpa \nabla \cdot \mpe -
          \mpe \cdot \nabla \mpa \right) \Big] \, dy \,
          dx. 
\end{align}
This formula forms the basis for all of the analysis throughout the
rest of the paper.

\section{Edge domain walls}
\label{sec:monodomain-states}

We next focus on the minimizers of $E_0$ from Eq.~\eqref{E0} in the
case of $w \gg 1$ and $\kappa$ below the threshold of the onset of
helicoidal structures corresponding to $x$-independent ground state
magnetization configurations. \cite{Vedmedenko2004,
  Pfleiderer04,Uchida06,Tretiakov_DMI} From the physical
considerations (for a rigorous mathematical justification in the case
$h = 0$, see Ref.~\onlinecite{Muratov2016}), it is clear that in these
states the magnetization vector will rotate in the $yz$-plane. Hence,
introducing the ansatz:
\begin{align}
  \label{mmono}
  \m(y) = (0, \sin \theta(y), \cos \theta(y)),
\end{align}
into Eq.~\eqref{E0}, we rewrite $E_0(\m)$ as
\begin{align}
  \label{E0mono}
  E_0(\m) = \int_{-w/2}^{w/2} \Big[ |\theta'|^2 + (Q - 1) \sin^2
  \theta \notag \\
  - 2 h \sin \theta + \kappa \theta' \Big] \, dy. 
\end{align}
The corresponding Euler-Lagrange equation associated with $E_0$ is
\begin{align}
  \label{E0monoEL}
  \theta'' - (Q - 1) \sin \theta \cos \theta + h \cos \theta = 0,
\end{align}
with boundary conditions
\begin{align}
  \label{E0monobc}
  \theta' \Big( \pm \frac{w}{2} \Big) =- {\kappa \over 2}.
\end{align}
Note that Eqs.~\eqref{E0monoEL} and \eqref{E0monobc} obey the
following symmetry relation, which leaves the energy $E_0$ unchanged:
\begin{align}
  \label{monosym}
  \theta \to \pi - \theta, \qquad y \to -y.
\end{align}
Introducing
\begin{align}
  \label{thetah}
  \theta_h = \arcsin \left( {h \over Q - 1} \right),
\end{align}
we first notice that when $w \to \infty$, we should have either
$\theta \to \theta_h$ or $\theta \to \pi - \theta_h$, corresponding to
the two monodomain ground states in the extended film for
$0 \leq h < Q - 1$. In view of the symmetry in
Eq.~\eqref{monosym}, it is enough to consider only the former
case.

In computing the minimal value $e_0(h, w)$ of $E_0$ for $w \gg 1$ one
needs to take into account the contributions of the boundary layers
next to $y = \pm \tfrac12 w$, the so-called edge domain
walls\cite{Muratov2016}. Below we show that in the presence of an
applied field the minimal energy admits an expansion of the
following form as $w \gg 1$:
\begin{align}
  e_0(h, w) & = -(Q - 1)^{-1} h^2 w \notag \\ 
            &\quad + \sigma_{edge}^+(h) +
              \sigma_{edge}^-(h) + g_0(h, w),
  \label{e0}
\end{align}
where the first term is the contribution of $\theta = \theta_h$ in the
bulk and, the second two terms are the edge domain wall contributions
from the upper and lower edge, respectively, whose explicit form will
be determined shortly, and the last term is an exponentially small
correction that is negligible for $w \gg 1$.

\begin{figure}[t]
  \centering
  \includegraphics[width=3.1in]{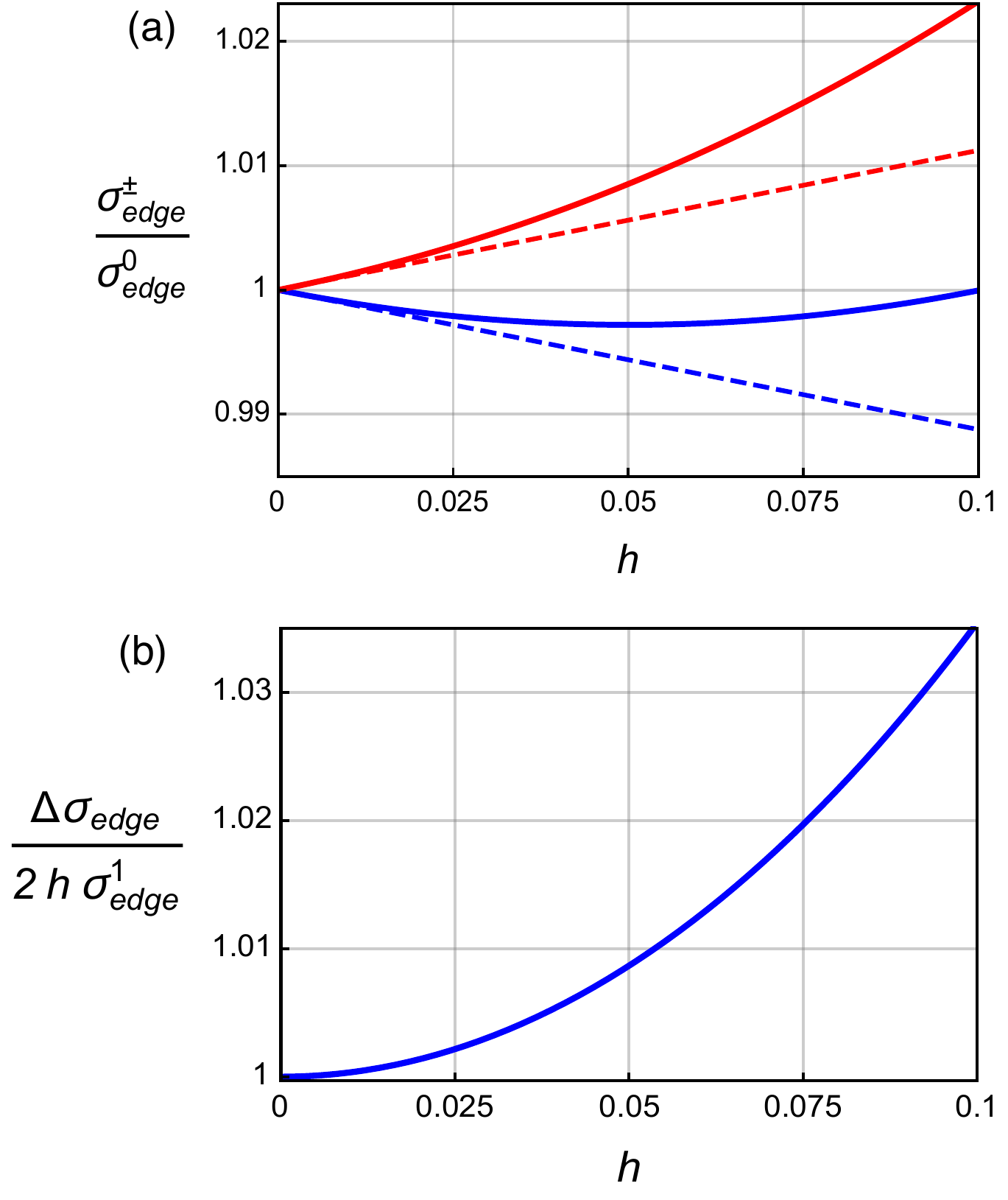} 
  \caption{Comparison of the exact edge domain wall energies in
    Eq.~\eqref{sedge} with the approximate ones given by
    Eq.~\eqref{sedgepm} for $Q = 1.674$ and $\kappa = 0.366$ (see
    Sec.~\ref{sec:comp-with-numer} for the corresponding material
    parameters. For those parameters, the value of $h = 0.1$
    corresponds to $\mu_0 H = 132$ mT). (a) Blue and red curves
    correspond to $\sigma_{edge}^+$ and $\sigma_{edge}^-$,
    respectively. Solid lines correspond to the exact values from
    Eq.~\eqref{sedge} and dashed lines represent the approximation of
    Eq.~\eqref{sedgepm}. (b) The relative error in approximating
    $\Delta \sigma_{edge} = \sigma_{edge}^+ - \sigma_{edge}^-$
    using Eq.~\eqref{sedgepm} is shown.}
  \label{fig:edgeexactvslin}
\end{figure}

We now derive Eq.~\eqref{e0}. Close to $y = \pm \tfrac12 w$ the
solutions of Eqs.~\eqref{E0monoEL} and \eqref{E0monobc} approaching
$\theta_h$ in the sample interior are expected to be well approximated
by those on half-line approaching $\theta_h$ far from the edge. After
a straightforward integration, we obtain
$\theta(y) \simeq \theta^\pm(y \mp \tfrac12 w)$,
where\cite{Goussev2013prb}
\begin{align}
  \label{thh}
  \theta^\pm(y) 
  & = 2 \tan ^{-1}\Bigg(\tan
    \left(\frac{\theta_h}{2}\right) \notag \\ 
  &  \quad +\frac{\cos (\theta_h) \sec  
    ^2\left(\frac{\theta_h}{2}\right)}{\tan
    \left(\frac{\theta_h}{2}\right) \mp e^{\mp\sqrt{Q-1} (y - y_0^\pm) 
    \cos (\theta_h)}}\Bigg).
\end{align}
The unknown values of $y_0^\pm$ are obtained by substituting the above
expression into Eq.~\eqref{E0monobc}, yielding
\begin{align}
  \label{ypm}
  y_0^\pm = \pm \frac{\cosh
  ^{-1}\left(\frac{2 \sqrt{Q-1} \cos ^2 \theta_h}{\kappa } \pm \sin
  \theta_h\right)}{\cos \theta_h 
  \sqrt{Q-1}}. 
\end{align}
Introducing $\theta_0^\pm = \theta^\pm(0)$, where, after simplifying
the obtained expressions, one gets explicitly
\begin{align}
  \label{th0pm}
  \theta_0^\pm = \arcsin \left( \sin \theta_h \mp {\kappa \over 2
  \sqrt{Q - 1}} \right).
\end{align}
We can then compute the contributions of the profiles in
Eq.~\eqref{thh} by plugging them into the energy in
Eq.~\eqref{E0mono}. After a rather tedious calculation, up to an
exponentially small error $g_0(h, w)$ we obtain Eq.~\eqref{e0} with
$\sigma_{edge}^\pm$ given explicitly by
\begin{align}
  \label{sedge}
  \sigma_{edge}^\pm 
  & = 2 \sqrt{Q - 1} \big( \theta_h \sin \theta_h +
    \cos \theta_h \notag \\
  & \quad - \cos \theta_0^\pm - \theta_0^\pm \sin \theta_h 
    \big) \pm \kappa (\theta_0^\pm - \theta_h).
\end{align}

Focusing on the regime of moderate values of $h \lesssim 1$, which is
the main regime of practical interest, linearizing Eq.~\eqref{sedge}
in $h$ we get
\begin{align}
  \label{sedgepm}
  \sigma_{edge}^\pm \simeq \sigma_{edge}^0 \pm \sigma_{edge}^1 h,
  \qquad 0 < h \lesssim 1,
\end{align}
where
\begin{align}
  \sigma_{edge}^0 & = 2 \sqrt{Q - 1} \left( 1 - \sqrt{1 - {\kappa^2
                    \over 4 (Q - 1)} } \, \right) \notag \\
                  & \quad \qquad \quad \qquad - \kappa \arcsin \left(
                    {\kappa  \over 2 \sqrt{Q - 1} }
                    \right), \label{sedgepm0} \\
  \sigma_{edge}^1 & = {2 \over \sqrt{Q - 1}} \arcsin \left( {\kappa
                    \over 2 \sqrt{Q - 1}} \right) - {\kappa \over Q -
                    1}. \label{sedgepm1}
\end{align}
We note that Eqs.~\eqref{sedgepm} gives a very good approximation to
the exact expression in Eq.~\eqref{sedge}, see
Fig. \ref{fig:edgeexactvslin}(a). In fact, the difference
$\sigma_{edge}^+ - \sigma_{edge}^-$, which is the relevant quantity
for the domain wall tilt, is captured by Eq.~\eqref{sedge} within a
few percent for practically all values of $h$ and $\kappa$, see
Fig. \ref{fig:edgeexactvslin}(b).

Before concluding this section, we make several observations regarding
Eq.~\eqref{sedgepm}. First, as expected, $\sigma_{edge}^{0,1} \to 0$
as $\kappa \to 0$, indicating that the edge domain walls disappear
without DMI irrespectively of the magnitude of $h$. Of course, the
same conclusion holds for Eq.~\eqref{sedge} as well. Second, for
$h \lesssim 1$ the applied field affects the contribution of the edge
domain walls to the energy only through $\sigma_{edge}^1$. At the same
time, it easy to see that as a function of $\kappa$ we have
$\sigma_{edge}^1 = O(\kappa^3)$, indicating that the effect of the
edge walls is negligible when the surface DMI is sufficiently weak.

\section{One-dimensional interior wall profile}
\label{sec:interior-wall}

We now turn to interior walls and obtain the leading order expressions
for the one-dimensional wall profiles and their energies as functions
of the wall orientation for $h \lesssim 1$. We focus mostly on the two
relevant cases that are amenable to an analytical treatment:
$h \ll \kappa \sim 1$ and $h \sim \kappa \ll 1$, even though our
method is applicable to all values of the parameters $\kappa$ and $h$
for which Dzyaloshinskii walls are expected to exist. For notational
convenience, we introduce the constant two-dimensional vector
\begin{align}
  \label{mpe0}
  \mpeb = (Q - 1)^{-1} h \hat{\mathbf y},
\end{align}
equal to the in-plane component of the equilibrium magnetization in
the film bulk.

We consider a one-dimensional profile in the direction
$\n_\alpha = (\cos \alpha, \sin \alpha)$, namely, a magnetization
configuration $\m(\xi) = (\mpe(\xi), \mpa(\xi))$, where
$\xi = \mathbf r \cdot \n_\alpha$. Then from Eq.~\eqref{E} with
$w = \infty$ the energy per unit length of the wall profile $\m(\xi)$
is
\begin{eqnarray}
  \label{Eal}
 \!\!\!\! \!\! E_\alpha(\m) \!\!&=& \!\!\int_{-\infty}^\infty \Big[ |\m'|^2 + (Q
  - 1) |\mpe - \mpeb|^2  \nonumber \\
  &&\! + \kappa \left( \mpa (\n_\alpha \cdot \mpe') - \mpa' (\mpe \cdot
  \n_\alpha) \right) \Big] d\xi.
\end{eqnarray}
The profile is to satisfy the following conditions at infinity:
\begin{align}
  \label{malinf}
  \mpe (\pm \infty) = \mpeb, \qquad \mpa(\pm
  \infty) = \pm \sqrt{1 - |\mpeb|^2},
\end{align}
and the associated Euler-Lagrange equation is
\begin{align}
  \mpe'' - (Q - 1) (\mpe - \mpeb) + \kappa \mpa' \n_\alpha 
  & =
    \lambda(\xi) \mpe , 
  \label{ELinperp}
  \\
  \mpa'' - \kappa (\n_\alpha \cdot \mpe')
  &  = \lambda(\xi) \mpa, \label{ELinpar} 
\end{align}
where $\lambda(\xi)$ is a scalar Lagrange multiplier due to the
pointwise unit length constraint on $\m$. The wall energy
$\sigma_{wall}$ associated with a solution $\m = \m^\alpha(\xi)$ of
Eqs.~\eqref{ELinperp} and~\eqref{ELinpar} satisfying \eqref{malinf} is
defined as
\begin{align}
  \label{sigwgen}
  \sigma_{wall}(\alpha) = E_\alpha(\m^\alpha).  
\end{align}
A distinctive feature of the wall energy in Eq.~\eqref{sigwgen} is
that for $\kappa \not= 0$ it depends on the wall orientation $\mathbf
n_\alpha$. 

It is not possible to find an analytical solution to the system of
Eqs.~\eqref{malinf}--\eqref{ELinpar} for general values of $Q$,
$\kappa$, $h$, and $\alpha$. Although it is not difficult to construct
such solutions numerically for any given set of the parameters (see
Sec.~\ref{sec:comp-with-numer}).

\subsection{$h \ll \kappa \sim 1$ regime}
\label{sec:regime1}

We now wish to obtain the leading order expansion of
$\sigma_{wall}(\alpha)$ for $h \ll 1$ and $\kappa \sim 1$.  Setting
$h = 0$ in Eqs.~\eqref{ELinperp} and \eqref{ELinpar} yields the
equation for the profile $\m_0 = \m_0(\xi)$:
\begin{align}
  \mpeO'' - (Q - 1) \mpeO  + \kappa \mpaO' \n_\alpha 
  & =
    \lambda_0(\xi) \mpeO, 
  \label{EELinperp}
  \\
  \mpaO'' - \kappa (\n_\alpha \cdot \mpeO')
  &  = \lambda_0(\xi) \mpaO. \label{EELinpar} 
\end{align}
The solution of Eqs.~\eqref{EELinperp} and \eqref{EELinpar} that
satisfies \eqref{malinf} is explicitly given by
\begin{align}
  \label{m0}
  \begin{split}
    \m_0(\xi) & = (\n_\alpha \sin \theta_0(\xi), \cos \theta_0(\xi)), \\
    \theta_0(\xi) & = 2 \arctan e^{-\xi \sqrt{Q - 1}},
  \end{split}
\end{align}
and for $h = 0$ we have $E_\alpha(\m_0) = \sigma_{wall}^0$, where
\begin{align}
  \label{sigw0}
    \sigma_{wall}^0 & = 4 \sqrt{Q - 1} - \pi \kappa.
\end{align}
Notice that $\sigma_{wall}^0$ does not depend on $\alpha$. 

To obtain the leading order correction to $\sigma_{wall}^0$, we write
$\m^\alpha = \m_0 + \m_1$, where $|\m_1| \ll |\m_0| = 1$, and note
that due to the pointwise unit length constraint, we have
$\m_0 \cdot \m_1 \simeq 0$ to the leading order. Next, we substitute
this expansion into Eq.~\eqref{Eal} to obtain, keeping only the terms
that are linear in $\m_1$ and $\mpeb$:
\begin{align}
  \label{EEal}
  \sigma_{wall}
  & (\alpha) 
    \simeq \sigma_{wall}^0 + \int_{-\infty}^\infty
    \Big[ 2 \m_0' \cdot \m_1' \notag \\
  & \,\,\qquad + 2 (Q-1) \mpeO \cdot (\mpeI - \mpeb) \notag \\  
  & \,\,\qquad+ \kappa \Big(\mpaO (\n_\alpha \cdot \mpeI') + \mpaI
    (\n_\alpha  \cdot \mpeO') \notag \\
  & \,\,\qquad- (\n_\alpha \cdot \mpeO) \mpaI' - (\n_\alpha \cdot \mpeI)
    \mpaO' \Big)  \Big] \, d\xi. 
\end{align}
Integrating by parts and using Eq.~\eqref{malinf}, this expression may
be rewritten equivalently as
\begin{align}
  \label{EEEal}
  & \sigma_{wall} (\alpha) 
    \simeq \sigma_{wall}^0 + 2 \kappa \n_\alpha \cdot \mpeb \notag \\
  & \quad - 2 (Q -
    1) \int_{-\infty}^\infty \mpeO  \cdot \mpeb d\xi \nonumber \\
  & \quad - 2 \int_{-\infty}^\infty
    \Big[ \m_0'' \cdot \m_1 - (Q-1) \mpeO \cdot \mpeI \notag \\
  & \quad - \kappa \Big( \mpaI (\n_\alpha
    \cdot \mpeO') -  (\n_\alpha \cdot \mpeI)
    \mpaO' \Big)  \Big] \, d\xi. 
\end{align}
In fact, in the above formula the integrand in the last integral is
zero to the leading order, which can be seen by multiplying both sides
of the Euler-Lagrange equation in Eqs.~\eqref{EELinperp} and
\eqref{EELinpar} by $\m_1$ and using the condition
$\m_0 \cdot \m_1 = 0$ to the leading order in $h$. Thus, substituting
the profile $\m_0$ into the above expression, after some more algebra
we get that the wall energy up to $O(h^2)$ is
\begin{align}
\label{Sigwall0}
  \sigma_{wall}(\alpha) \simeq \sigma_{wall}^0 - \sigma_{wall}^1 h \sin
  \alpha,
\end{align}
where
\begin{align}
\label{Sigwall1}
  \sigma_{wall}^1 & = \frac{2\pi}{\sqrt{Q-1}} - \frac{2 \kappa}{Q-1}. 
\end{align}

We point out that the obtained expression for $\sigma_{wall}(\alpha)$
appears to be meaningless when $\kappa =0$, since Eq.~\eqref{Sigwall1}
suggests that for $h > 0$ the wall energy depends on the angle
$\alpha$ even in the absence of DMI. Yet the energy in Eq.~\eqref{Eal}
is manifestly independent of $\alpha$. The reason for this discrepancy
is the fact that our approximations are justified only when
$\kappa \sim 1 \gg h$, while the limit of $\kappa \to 0$ with $h > 0$
fixed violates this assumption. In fact, when $\kappa \sim 1$ the
magnetization in a domain wall rotates mostly in the plane spanned by
$\n_\alpha$ and $\hat{\mathbf z}$, while when $\kappa = 0$ the
magnetization would prefer to rotate in the plane spanned by
$\hat{\mathbf y}$ and $\hat{\mathbf z}$, even if
$\n_\alpha \not= \hat{\mathbf y}$. To resolve this discrepancy, we
need to consider the case of $\kappa \lesssim h \ll 1$ separately.

\subsection{$h \sim \kappa \ll 1$ regime}
\label{sec:regime2}

When both $h$ and $\kappa$ are small and comparable, we can further
simplify the argument above to obtain the following equation for
$\m_0$ in place of Eqs.~\eqref{EELinperp} and \eqref{EELinpar} to the
leading order:
\begin{align}
  \mpeO'' - (Q - 1) \mpeO  
  & =
    \lambda_0(\xi) \mpeO, 
  \label{EEELinperp}
  \\
  \mpaO'' 
  &  = \lambda_0(\xi) \mpaO. \label{EEELinpar} 
\end{align}
The solution of Eqs.~\eqref{EEELinperp} and \eqref{EEELinpar} that
satisfies \eqref{malinf} is explicitly given by $\m_0 =
\m_0^\mathbf{n}$, where
\begin{align}
  \label{m0u}
  \begin{split}
    \m_0^\mathbf{n}(\xi)  & = (\mathbf n \sin \theta_0, \cos \theta_0), \\
    \theta_0(\xi) & = 2 \arctan e^{-\xi \sqrt{Q - 1}},
  \end{split}
\end{align}
and $\mathbf n \in \R^2$ is an arbitrary constant unit vector. For
$h = \kappa = 0$ we have
$E_\alpha(\m_0^\mathbf{n}) = \sigma_{wall}^0$, where
\begin{align}
  \label{sigw0bis}
  \sigma_{wall}^0 & = 4 \sqrt{Q - 1}.
\end{align}
Notice that $\sigma_{wall}^0$ does not depend on $\alpha$ or
$\mathbf n$ and coincides with the energy of the N\'eel wall in the
absence of nonlocal effects.

Now, writing again $\m^\alpha = \m_0^\mathbf{n} + \m_1$ and expanding
the energy to the next order in $h$ and $\kappa$, we obtain
\begin{eqnarray}
  \label{Ealcbis}
\!\!  E_\alpha (\m) 
            \!\!&\simeq \!\! & \sigma_{wall}^0 + \int_{-\infty}^\infty
            \Big[ 2 \m_0' \cdot \m_1' \nonumber \\
            &&+ \kappa  \left( \mpaO (\n_\alpha
            \cdot \mpeO') - \mpaO' (\mpeO \cdot 
            \n_\alpha) \right)   \nonumber \\
          && + 2 (Q-1) \mpeO \cdot (\mpeI - \mpeb)  \Big] d\xi,
\end{eqnarray}
and following the same arguments as in Sec.~\ref{sec:regime1} we
arrive at
\begin{align}
  \label{Ealcubis}
  E_\alpha(\m) = \sigma_{wall}^0 - \frac{2\pi h}{\sqrt{Q-1}} ({\mathbf n}
  \cdot  \hat{\mathbf y}) - \pi \kappa ( {\mathbf n} \cdot \n_\alpha). 
\end{align}

Finally, in order to find the direction of vector $\mathbf n$ we need
to minimize the above energy with respect to $\mathbf n$. It is easy
to see that
\begin{align} 
  {\mathbf n} = \frac{2\pi h
  \hat{\mathbf y} + \pi \kappa \n_\alpha \sqrt{Q-1}}{\left| 2\pi
  h \hat{\mathbf y} + \pi \kappa \n_\alpha \sqrt{Q - 1} \right|
  }
\end{align}
minimizes the right-hand side of Eq.~\eqref{Ealcubis}, and the minimum
of the energy is given by
\begin{eqnarray}
  \label{swall2}
 \!\!\!\!\!\!\!\! \sigma_{wall}(\alpha) \!\!& \simeq \!\!& 4\sqrt{Q-1} \nonumber \\
  && - \pi \sqrt{ \kappa^2 \cos^2\alpha + 
    \left(\kappa\sin\alpha + \frac{2 h}{\sqrt{Q-1}}\right)^2}. 
\end{eqnarray}

Thus, the obtained magnetization profile rotates mostly in the plane
spanned by $\mathbf n$ and $\hat{\mathbf z}$, with $\mathbf n$
depending sensitively on both $h$ and $\kappa$. Furthermore, the
obtained result is consistent with the one of
Sec. \ref{sec:regime1}. Indeed, expanding the expression in
Eq.~\eqref{swall2} in the powers of $h$ with $\kappa \ll 1$ fixed
yields Eq.~\eqref{Sigwall0} to linear order in $h$ and the leading
order in $\kappa$. At the same time, setting $\kappa = 0$ with
$0 < h \ll 1$ fixed in Eq.~\eqref{Ealcubis}, we recover the wall
energy
$\sigma_{wall} \simeq 4 \sqrt{Q - 1} - {2 \pi h \over \sqrt{Q - 1}}$,
which is easily seen to be the wall energy for a profile rotating in
the plane spanned by $\hat{\mathbf y}$ and $\hat{\mathbf z}$,
consistent with the discussion at the end of Sec.~\ref{sec:regime1}.

\section{Two-dimensional problem}
\label{sec:two-dimens-probl}

\begin{figure}[t]
  \centering
  \includegraphics[width=3.15in]{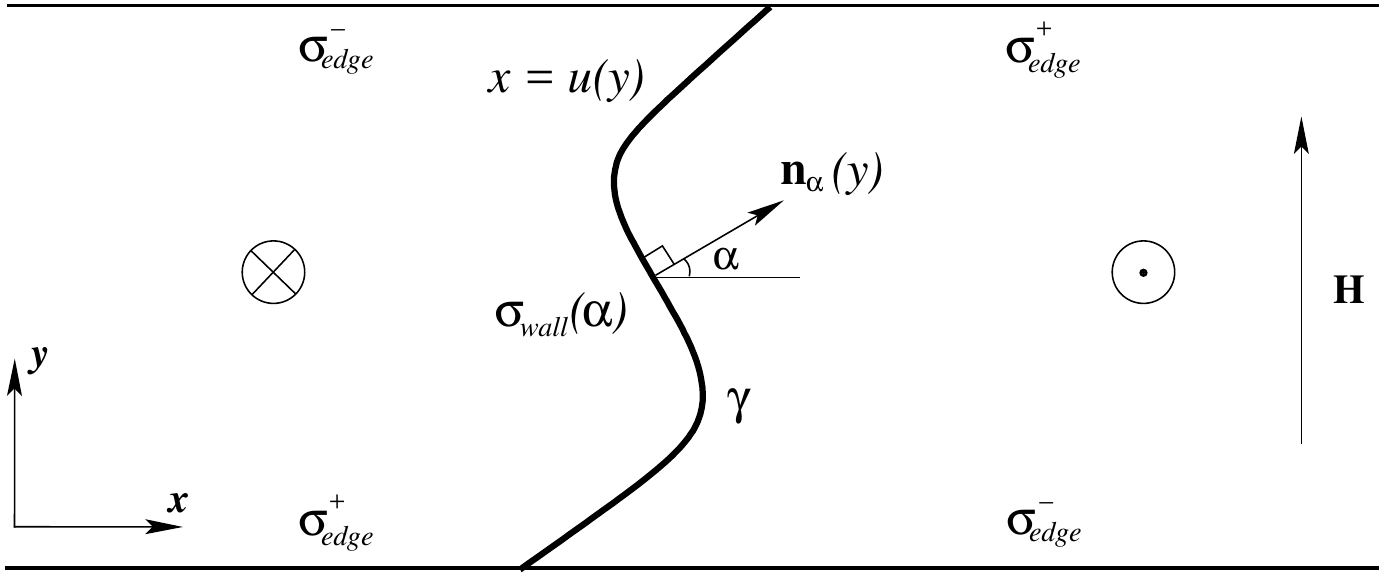}
  \caption{Schematics of a general domain wall geometry in an infinite
    strip. The up/down symbols indicate the direction of the
    magnetization far from the wall and the strip edges.}
  \label{fig:twodim}
\end{figure}

We now demonstrate how the information about one-dimensional domain
walls obtained in the preceding sections may be applied to a single
Dzyaloshinskii domain wall running across an infinite ferromagnetic
nanostrip. For an illustration of the geometry, see
Fig.~\ref{fig:twodim}, where the domain wall is represented by a thick
solid curve. Here we wish to treat the wall as a one-dimensional
object, whose shape is determined by minimizing an appropriate
geometric energy functional. This energy functional is obtained via a
suitable asymptotic reduction of the two-dimensional micromagnetic
energy in Eq.~\eqref{E}. For a rigorous justification of such an
approach in a closely related context, see
Ref.~\onlinecite{Muratov2016}.

Using Eq.~\eqref{e0}, we can rewrite Eq.~\eqref{E} in the following
way: 
\begin{align}
  \label{EE}
  E(\m) & = \int_{-\infty}^\infty \int_{-w/2}^{w/2} \Big[ |\nabla \m|^2 + (Q
          - 1) |\mpe - \overline{\m}_\perp|^2 \notag \\
        & - w^{-1} (\sigma_{edge}^+ +
          \sigma_{edge}^-) - w^{-1} g_0(h, w) \notag \\
        & +  \kappa \left( \mpa \nabla \cdot \mpe - 
          \mpe \cdot \nabla \mpa \right) \Big] \, d^2 r.
\end{align}
Recall that $\overline{\m}_\perp$ was defined in Eq.~\eqref{mpe0}.  We
next consider a domain wall whose shape is described by a smooth curve
$\gamma$ which is the graph of a function
$u: (-w/2, w/2) \to \R$, i.e., for every $\mathbf r \in \gamma$ we
have $\mathbf r = (u(y), y)$ for some $y \in (-w/2, w/2)$. The
associated magnetization profile ${\mathfrak m}_\gamma$ in the
vicinity of this curve will then be close to the optimal
one-dimensional interior wall profile analyzed in
Sec. \ref{sec:interior-wall}. Let $\mathbf r$ be a point in the
vicinity of $\gamma$ and let $\mathbf r_\gamma$ be the orthogonal
projection of $\mathbf r$ on $\gamma$. Denote by
$\mathbf n_\alpha(\mathbf r_\gamma) = (\cos \alpha(\mathbf r_\gamma),
\sin \alpha(\mathbf r_\gamma))$
the unit normal vector to $\gamma$ at point $\mathbf r_\gamma$
pointing towards the region where $\mpa > 0$, with
$\alpha(\mathbf r_\gamma)$ the angle that the normal vector
$\mathbf n_\alpha(\mathbf r_\gamma)$ makes with the $x$-axis at
point $\mathbf r_\gamma$.  Then the magnetization profile
$\m = {\mathfrak m}_\gamma$ associated with the curve $\gamma$ is
expected to satisfy
\begin{align}
  \label{ansint}
  {\mathfrak m}_\gamma(\mathbf r) \simeq \m^\alpha ((\mathbf r -
  \mathbf r_\gamma) \cdot \mathbf n_\alpha(\mathbf r_\gamma)),
\end{align}
where $\m^\alpha$ is the optimal profile that minimizes the
one-dimensional interior wall energy $E_\alpha$ in
Eq.~\eqref{Eal}. Assuming that the curvature of $\gamma$ does not
exceed $O(w^{-1})$, the contribution to the energy by the neighborhood
of $\gamma$ for $w \gg 1$ is then dominated by the one-dimensional
wall energy $E_\alpha(\m^\alpha)$ integrated over $\gamma$, which by
Eq.~\eqref{sigwgen} is
\begin{align}
  E_{int}({\mathfrak m}_\gamma) = \int_\gamma
  \sigma_{wall}(\alpha(\mathbf r)) ds(\mathbf r),
\end{align}
where $ds$ is the arclength differential along $\gamma$. Thus, the
energy of the interior wall is characterized by an \emph{anisotropic}
line tension.

Away from the interior wall the magnetization obeys 
\begin{align}
  \mathfrak m_\gamma \simeq \left( \overline{\m}_\perp, \pm \sqrt{1 - 
  |\overline{\m}_\perp|^2} \right),   
\end{align}
consistently with Eq.~\eqref{ansint}. However, this relation is
violated close to the strip edges, where edge domain walls
appear. Therefore, we also need to take into account the edge domain
wall profiles analyzed in Sec.~\ref{sec:monodomain-states} in those
regions. Accordingly, one expects
\begin{align}
  \mathfrak m_\gamma \simeq (0, \sin (\theta^+(y - \tfrac12 w)),
  \cos (\theta^+(y - \tfrac12 w) )), 
\end{align}
for $x > u(w/2)$ and $y \sim w/2$, whereas
\begin{align}
  \mathfrak m_\gamma \simeq (0, \sin (\theta^-(y + \tfrac12 w)),
  \cos (\theta^-(y + \tfrac12 w) )), 
\end{align}
for $x > u(-w/2)$ and $y \sim -w/2$. Similarly, in view of the
symmetry relation given by Eq.~\eqref{monosym}, we also find
\begin{align}
  \mathfrak m_\gamma \simeq (0, \sin (\theta^-(\tfrac12 w - y)),
  \cos (\theta^-(\tfrac12 w - y) )), 
\end{align}
for $x < u(w/2)$ and $y \sim w/2$, whereas
\begin{align}
  \mathfrak m_\gamma \simeq (0, \sin (\theta^+(-\tfrac12 w - y)),
  \cos (\theta^+(-\tfrac12 w - y) )), 
\end{align}
for $x < u(-w/2)$ and $y \sim -w/2$. The corresponding edge wall
energy is then
\begin{align}
  E_{edge} (\mathfrak m_\gamma) = (\sigma_{edge}^+ - \sigma_{edge}^-)
  (u(-w/2) - u(w/2)),
\end{align}
recalling that we subtracted the contribution of
$\sigma_{edge}^+ + \sigma_{edge}^-$ in Eq.~\eqref{EE}. Finally, to
match the interior and the edge wall profiles near points
$x = u(\pm \tfrac12 w)$ and $y = \pm \tfrac12 w$, one uses the
construction from Ref.~\onlinecite{Muratov2016}, which can be seen
not to contribute to the energy to the leading order.

Putting all the leading order contributions to the energy in
Eq.~\eqref{EE} together, we obtain
\begin{align}
  E(\mathfrak m_\gamma) \simeq
  E_{int} (\mathfrak m_\gamma) + E_{edge}(\mathfrak m_\gamma). 
\end{align}
Then, using the parametrization $x = u(y)$ of the curve $\gamma$,
we find explicitly
\begin{align}
  \label{Egeom0}
  E(\mathfrak m_\gamma) 
  & \simeq \int_{-w/2}^{w/2} \sigma_{wall}\big( -
    \arctan u'(y) \big)  \sqrt{1 + |u'(y)|^2} \, dy \notag \\ 
  & \quad - ( \sigma_{edge}^+ - \sigma_{edge}^-) (u(w/2) - u(-w/2)), 
\end{align}
where we recall that $\alpha(\mathbf r) = -\arctan u'(y)$. 

As is well known\cite{Herring52} and can be easily seen directly from
Eq.~\eqref{Egeom0}, every critical point $\gamma$ of
$\gamma \mapsto E(\mathfrak m_\gamma)$ is a straight line. In
particular, minimizers of $E(\mathfrak m_\gamma)$ are straight domain
walls running across the strip. Thus, the only free parameter in the
problem is the difference between the $x$-positions $u(w/2) - u(-w/2)$
of the wall at the top and bottom edges. In fact, from the dimensional
considerations this difference is proportional to $w$, i.e., $w$ can
be scaled out of the energy. Thus, the only
free parameter of the minimization problem for $E(\mathfrak m_\gamma)$
is the tilt angle $\beta \in (-{\pi \over 2}, {\pi \over 2})$ that the
line $\gamma = \gamma_\beta$ makes with the $y$-axis. Note that this
angle coincides with the angle $\alpha$ defining the normal vector
$\mathbf n_\alpha$ of $\gamma_\beta$. To compute the tilt angle, we
substitute the straight line ansatz $\gamma_\beta$ into
Eq.~\eqref{Egeom0}, and the angle is then obtained by minimizing the
expression
\begin{align}
  \label{Emb}
  {E(\mathfrak m_{\gamma_\beta}) \over w} \simeq {\sigma_{wall}(\beta)
  \over \cos \beta} + (\sigma_{edge}^+ - \sigma_{edge}^-) \tan \beta.
\end{align}
over $\beta$, which completely characterizes existence and
multiplicity of tilted domain walls in the presence of DMI. It is
clear from Eq.~\eqref{Emb} that the equilibrium tilt angle is
independent of the strip width and depends only on the dimensionless
material parameters $\kappa$ and $Q$ and the dimensionless applied
field strength $h$. In fact, dimensional analysis shows that the
equilibrium tilt angle depends on these parameters only via two
combinations, $\kappa / \sqrt{Q - 1}$ and $h/(Q - 1)$.

To conclude this section, we note that, as expected, the tilt angle
becomes zero when the effect of the DMI vanishes. This can be readily
seen from Eq.~\eqref{Emb}, taking into account that for $\kappa = 0$
we have $\sigma_{edge}^+ = \sigma_{edge}^-$ and $\sigma_{wall}$
becomes independent of $\beta$, see Eqs.~\eqref{sedge} and
\eqref{Eal}. 

In the rest of this section, we consider two parameter regimes based
on analytical results presented in Sec. \ref{sec:regime1} and
\ref{sec:regime2} for which explicit expressions for the tilt can be
obtained. 

\subsection{$h \ll \kappa \sim 1$ regime}
\label{sec2:regime1}

In this regime, an approximate expression for $\sigma_{wall}(\alpha)$
is given by Eqs.~\eqref{sigw0}, \eqref{Sigwall0} and \eqref{Sigwall1},
and $\sigma_{edge}^\pm$ are given by
Eqs.~\eqref{sedgepm}--\eqref{sedgepm1}. Substituting these expressions
into Eq.~\eqref{Emb}, we obtain
\begin{align}
  {E(\mathfrak m_{\gamma_\beta}) \over w} \simeq {\sigma_{wall}^0
  \over \cos \beta} + h (2 \sigma_{edge}^1 - \sigma_{wall}^1) \tan
  \beta.  
\end{align}
Minimizing this expression yields the unique equilibrium tilt angle
\begin{align}
  \label{titlta}
  \beta = \arcsin \left( { \sigma_{wall}^1  - 2 \sigma_{edge}^1 \over
  \sigma_{wall}^0} \, h \right).
\end{align}
In particular, since we are in the regime of small applied fields the
equilibrium tilt angle is linear in $h$:
\begin{align}
  \label{titltalin}
  \beta \simeq  { 4 h \arccos \left( {\kappa \over 2 \sqrt{Q - 1}}
  \right) \over 4 (Q - 1) - \pi \kappa \sqrt{Q - 1}}.
\end{align}
This formula is one of the main findings of our paper. 

We note that the expression in Eq.~\eqref{titlta} formally coincides
with the formula for the contact angle of a triple junction between
three distinct phases\cite{Landau5}. Nevertheless, in addition to the
contribution of the difference of line tensions
$\sigma_{edge}^0 \pm \sigma_{edge}^1 h$ associated with the two edges,
the formula also contains a contribution $\sigma_{wall}^1$ due to
anisotropy of the line tension of Dzyaloshinskii wall.

\subsection{$h \sim \kappa \ll 1$ regime}
\label{sec2:regime2}

In this regime, the explicit expressions for $\sigma_{wall}(\alpha)$
is given by Eq.~\eqref{swall2}. At the same time, recalling that
the expression for $\sigma_{edge}^\pm$ in Eq.~\eqref{sedgepm} remains
valid also for $\kappa \ll 1$ and that $\sigma_{edge}^1 =
O(\kappa^3)$, one can see that the contribution of $\sigma_{edge}^+ -
\sigma_{edge}^-$ in Eq.~\eqref{Emb} is negligible. Thus, to the
leading order we arrive at
\begin{align}
  {E(\mathfrak m_{\gamma_\beta}) \over w} 
  & \simeq 
    \frac{4\sqrt{Q-1}}{\cos\beta} \notag \\
  & \quad - \pi \sqrt{ \kappa^2 +
    \left(\kappa\tan\beta + \frac{2 h}{\sqrt{Q-1} \,
    \cos\beta}\right)^2}.  \label{Embbis}
\end{align}
Note that the second term in Eq.~\eqref{Embbis} is a small
perturbation for the first term, which is a convex even function of
$\beta$ approaching infinity as $\beta \to \pm {\pi \over
  2}$. Therefore, the minimum in Eq.~\eqref{Embbis} is attained for
$|\beta| \ll 1$. 

To proceed further, we expand the right-hand side of
Eq.~\eqref{Embbis} in a Taylor series in $\beta$ up to second order
and keep only the leading terms in $h$ and $\kappa$. The result is
\begin{align}
  {E(\mathfrak m_{\gamma_\beta}) \over w} 
  & \simeq 
    4\sqrt{Q-1} - {2 \pi h \kappa \beta \over \sqrt{4 h^2 + \kappa^2
    (Q - 1)}} \notag \\
  & \quad + \beta^2 \left( 2 \sqrt{Q - 1} - \frac{\pi \kappa}{2} \right).  
  \label{Embbisbis}
\end{align}
Minimizing this expression in $\beta$ yields the equilibrium tilt
angle 
\begin{align}
  \label{betabis}
  \beta \simeq \frac{\pi  h \kappa }{\left(2 \sqrt{Q-1}-\frac{\pi
  \kappa }{2}\right) \sqrt{4 
  h^2+\kappa ^2 (Q-1)}}. 
\end{align}
This formula is another main finding of our paper. As expected, the
title angle in Eq.~\eqref{betabis} goes to zero as $h \to
0$.
Moreover, for $h \ll \kappa \ll 1$ we obtain an interesting result:
\begin{align}
  \label{betabis1}
  \beta \simeq {\pi h \over 2 (Q - 1)}, \qquad h \ll \kappa,
\end{align}
i.e., the equilibrium tilt angle becomes {\em independent} of the DMI
strength. In fact, this is in agreement with the prediction of
Eq.~\eqref{titltalin} for vanishingly small $\kappa$.

Similarly, when $\kappa \ll h \ll 1$, we find another surprising
result: 
\begin{align}
  \label{betabis2}
  \beta \simeq {\pi \kappa \over 4 \sqrt{Q - 1}}, \qquad \kappa \ll h,
\end{align}
i.e., the equilibrium tilt angle becomes {\em independent} of the
applied field. This indicates that for moderate values of the DMI
strength the measured tilt angle may be used to directly assess the
value of the interfacial DMI constant experimentally.

\section{Comparison with micromagnetic simulations}
\label{sec:comp-with-numer}

To validate the conclusions of our analysis, we performed three types
of numerical tests. For the material parameters, we chose those of
a 0.6 nm-thick film corresponding roughly to two monolayers
of Co, with parameters $A = 10^{-11}$J/m, $K = 1.25 \times 10^6$
J/m$^3$, $M_s = 1.09 \times 10^6$ A/m. The representative values of
the DMI strength and applied field are $D = 1$ mJ/m$^2$ and
$\mu_0 H = 100$ mT, respectively\cite{Boulle13}.

We begin by comparing the tilted Dzyaloshinskii domain wall profiles
from the two-dimensional numerical simulations 
obtained using {\sc Mumax3} simulation package within the local
approximation of the magnetostatic energy \footnote{This amounts to
  combining the anisotropy and magnetostatic energy contributions into
  a single effective anisotropy term with constant
  $K_{\rm{eff}}= K -\tfrac{1}{2} \mu_0 M_{s}^{2} =0.504\times 10^6$
  J/m$^3$ and neglecting the rest of dipolar effects.}  [as in
Eq.~\eqref{E}], with the 1D domain wall profiles $\m^\alpha$
minimizing $E_\alpha$ in Eq.~\eqref{Eal}. In the micromagnetic
simulations, we used a conservative discretization step of $1$ nm in
the $xy$-plane. To obtain the one-dimensional profiles $\m^\alpha$
minimizing $E_\alpha$, we solved Eqs.~\eqref{malinf}--\eqref{ELinpar}
by writing $\m^\alpha$ in polar coordinates for $\theta$ and $\phi$:
\begin{align}
  \label{mal}
  \m^\alpha = (\sin \theta \cos \phi, \sin \theta \sin \phi, \cos
  \theta), 
\end{align}
and solving the following evolution problem:
\begin{align}
  \theta_t & = \theta_{\xi\xi} - \left(\phi_\xi^2+Q-1\right) \sin
             \theta \cos \theta + h \cos \theta \sin \phi \notag \\
           & \quad -\kappa
             \phi_\xi \sin(\phi - \alpha) \sin ^2 \theta , \label{th} \\ 
  \phi_t & = \phi_{\xi\xi} + 2 \theta_\xi \phi_\xi \cot \theta + h
           \csc \theta \cos \phi \notag \\ 
           & \quad +\kappa \theta_\xi \sin (\phi - \alpha
             ), \label{phi} 
\end{align}
until a steady state was reached. Here the subscripts stand for
 the respective partial derivatives. The equations above correspond to an
overdamped Landau-Lifshitz-Gilbert equation, and their steady states
solve Eqs.~\eqref{ELinperp} and \eqref{ELinperp} upon substitution
into Eq.~\eqref{mal}. Also, in terms of $\theta$ and $\phi$ the wall
energy is
\begin{align}
  E_\alpha(\m^\alpha) & = \int_{-\infty}^\infty \bigg( \theta_\xi^2 +
                        \phi_\xi^2 \sin^2 \theta  + (Q - 1) \sin^2 \theta \notag \\
                      & \quad - 2 h \sin \theta
                        \sin \phi + {h^2 \over Q - 1} + \kappa
                        \theta_\xi \cos (\phi - \alpha) \notag \\
                      & \quad - \kappa \phi_\xi \sin (\phi - \alpha) \cos
                        \theta \sin \theta \bigg) d \xi. 
\end{align}

\begin{figure}[t]
  \centering
  \includegraphics[width=3.15in]{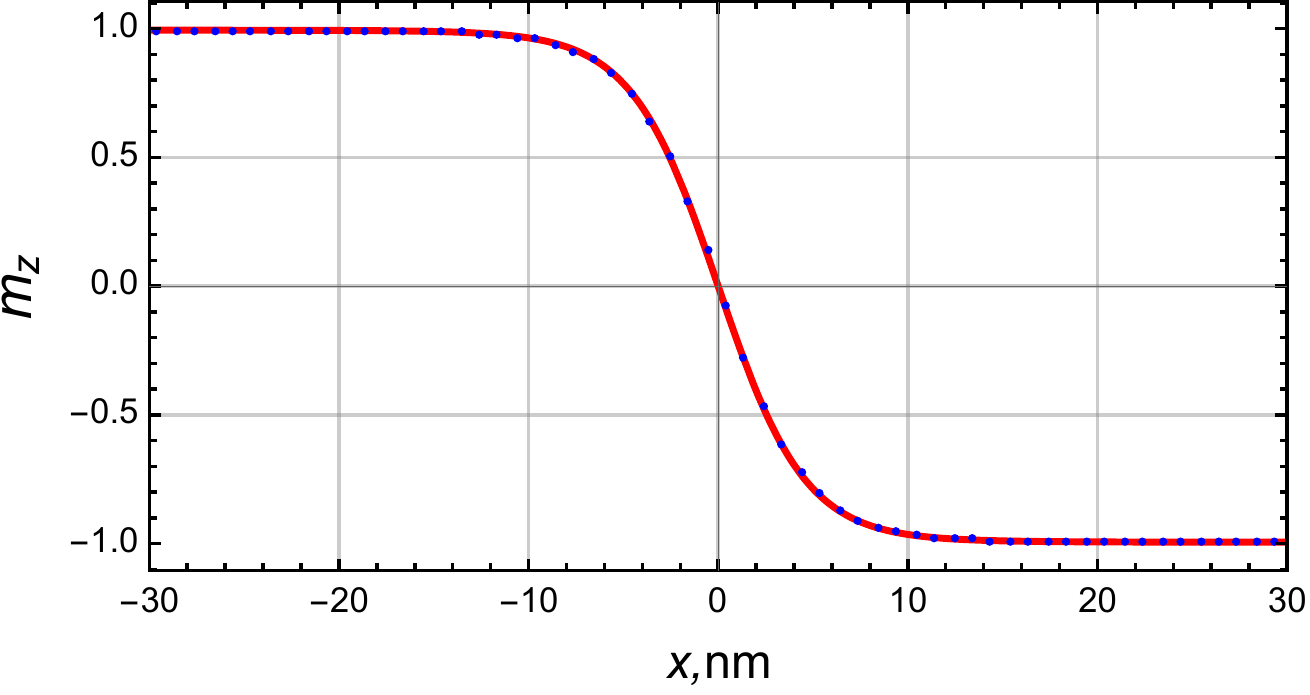}
  \caption{A one-dimensional $y = 0$ cut through the computed
    two-dimensional profile $\m$ (blue dots) vs. a
    one-dimensional cut through the optimal profile
    $\mathfrak m_{\gamma_\alpha}$ (red line). See text for
    details. }
  \label{fig:numprofiles}
\end{figure}

The parameters at the beginning of this section correspond to the
dimensionless parameters $Q = 1.674$, $\kappa = 0.366$ and
$h = 0.073$. For these parameters, we carried out {\sc
    Mumax3}\cite{Mumax} simulations in an 800 nm $\times$ 400 nm
strip, which corresponds to $w = 109 \gg 1$, and obtained the
magnetization profile with the tilt angle $\beta \simeq 11.2^\circ$.
We then solved Eqs.~\eqref{th} and \eqref{phi} with
$\alpha = 11.2^\circ$ and obtained the optimal one-dimensional
wall profile $\m^\alpha$. The result of the two-dimensional
computation is compared with the one-dimensional profile in
Fig.~\ref{fig:numprofiles}, which plots the $z$-component of the
two-dimensional profile $\m$ along the $x$-axis alongside with the
corresponding section of the optimal profile
$\mathfrak m_{\gamma_\alpha}$ obtained from $\m^\alpha$. One can see
an almost perfect agreement between the full two-dimensional
simulation result and the theoretical prediction of
Sec.~\ref{sec:interior-wall}. The same agreement is also observed in
the other two components of the magnetization (not shown). This
justifies the main premise of our theory about the one-dimensional
character of the interior wall profiles.

\begin{figure}[t]
  \centering
  \includegraphics[width=3.15in]{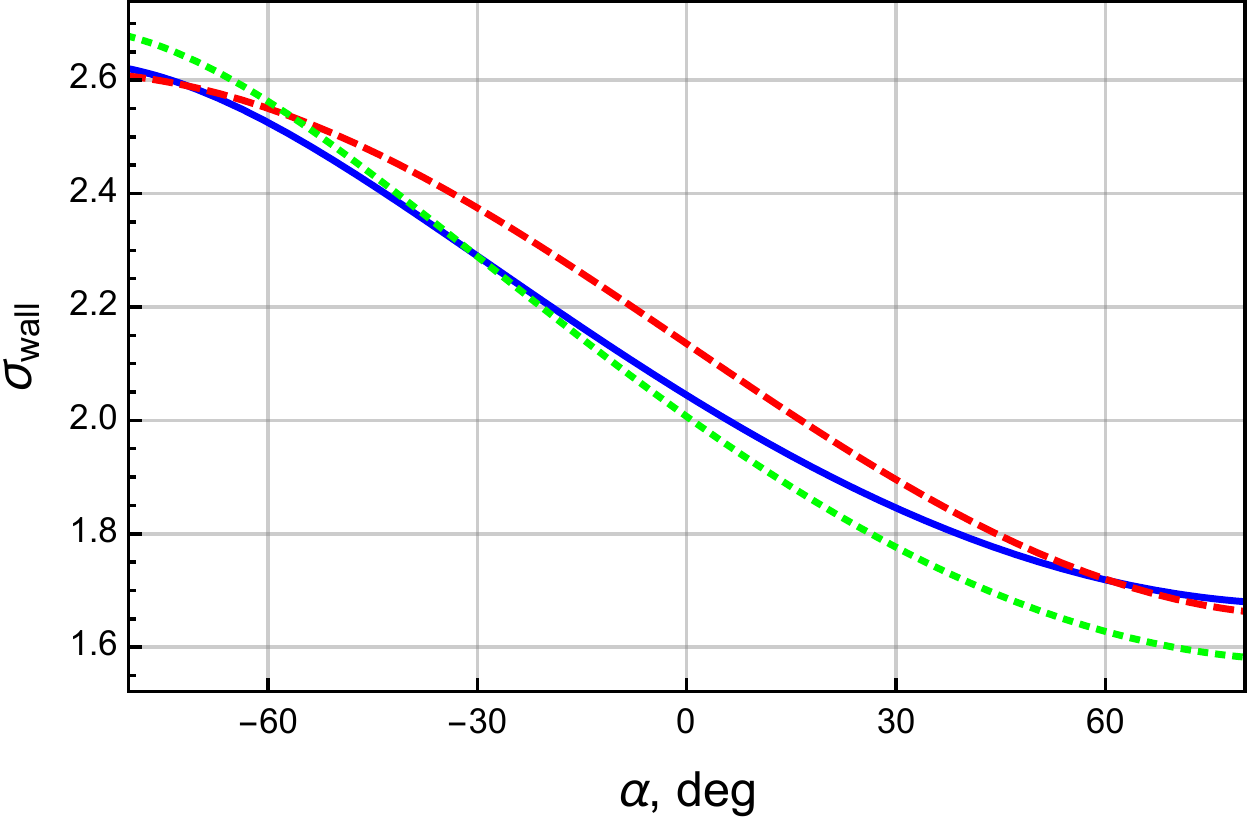}
  \caption{The dependence $\sigma_{wall}(\alpha)$ obtained from the
    numerical minimization of $E_\alpha$ (blue solid curve), the
    analytical expressions given by Eq.~\eqref{Sigwall0} (red dashed
    curve) and Eq.~\eqref{swall2} (green dotted curve),
    corresponding to the dimensionless parameters $Q = 1.674$,
    $\kappa = 0.366$ and $h = 0.073$.}
  \label{fig:numsigmawall}
\end{figure}

To further test the conclusions of our theory, we computed the energy
$\sigma_{wall}(\alpha)$ of the interior walls as a function of their
orientation angle $\alpha$ from the solutions of Eqs.~\eqref{th} and
\eqref{phi} for the considered values of the parameters. The result is
plotted in Fig.~\ref{fig:numsigmawall}, along with the analytical
approximations given by Eqs.~\eqref{Sigwall0} and \eqref{swall2}. One
can see that both analytical formulas give a fairly good approximation
to the exact interior wall energy $\sigma_{wall}(\alpha)$ for these
parameters. The agreement becomes much better for smaller values of
$h$. 

We used the interior wall energy $\sigma_{wall}(\alpha)$ obtained
numerically to calculate the equilibrium tilt angle by minimizing the
energy in Eq.~\eqref{Emb} numerically. This resulted in a unique
minimizing angle $\beta = 11.4^\circ$, in excellent agreement with the
result of the full two-dimensional simulation. For comparison, the
formulas in Eqs.~\eqref{titltalin} and \eqref{betabis} yield $\beta =
12.8^\circ$ and $\beta = 13.5^\circ$, respectively, still in a
good agreement with the two-dimensional result, which is reasonable
since both these formulas are at the limits of their applicability for
the considered parameters. 

\begin{figure}[t]
  \centering
  \includegraphics[width=3.15in]{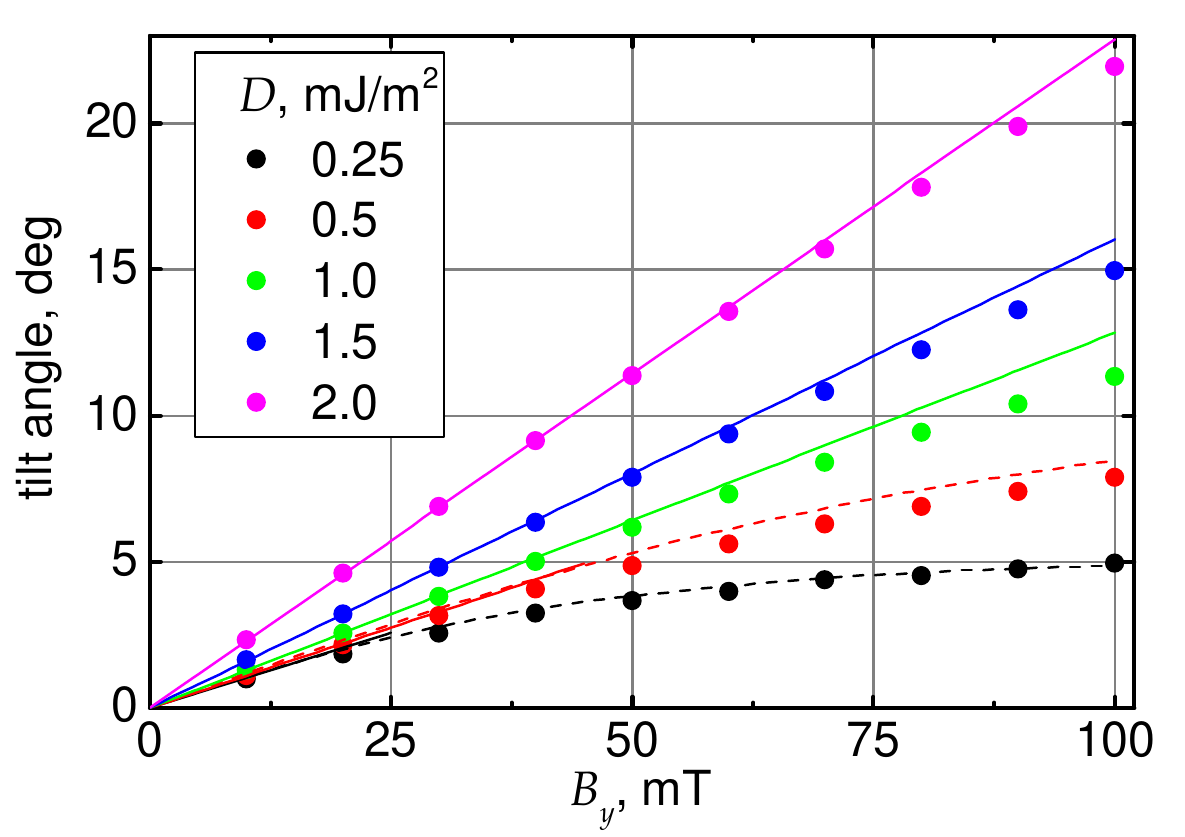}
  \caption{Equilibrium tilt angle, $\beta$, vs. applied field
    $B_y = \mu_0 H$ for several values of the DMI strength $D$,
    alongside with the predictions of Eq.~\eqref{titltalin} (solid
    lines) and Eq.~\eqref{betabis} (dashed lines).}
  \label{fig:tiltvsfield}
\end{figure}

\begin{figure}[t]
  \centering
  \includegraphics[width=3.15in]{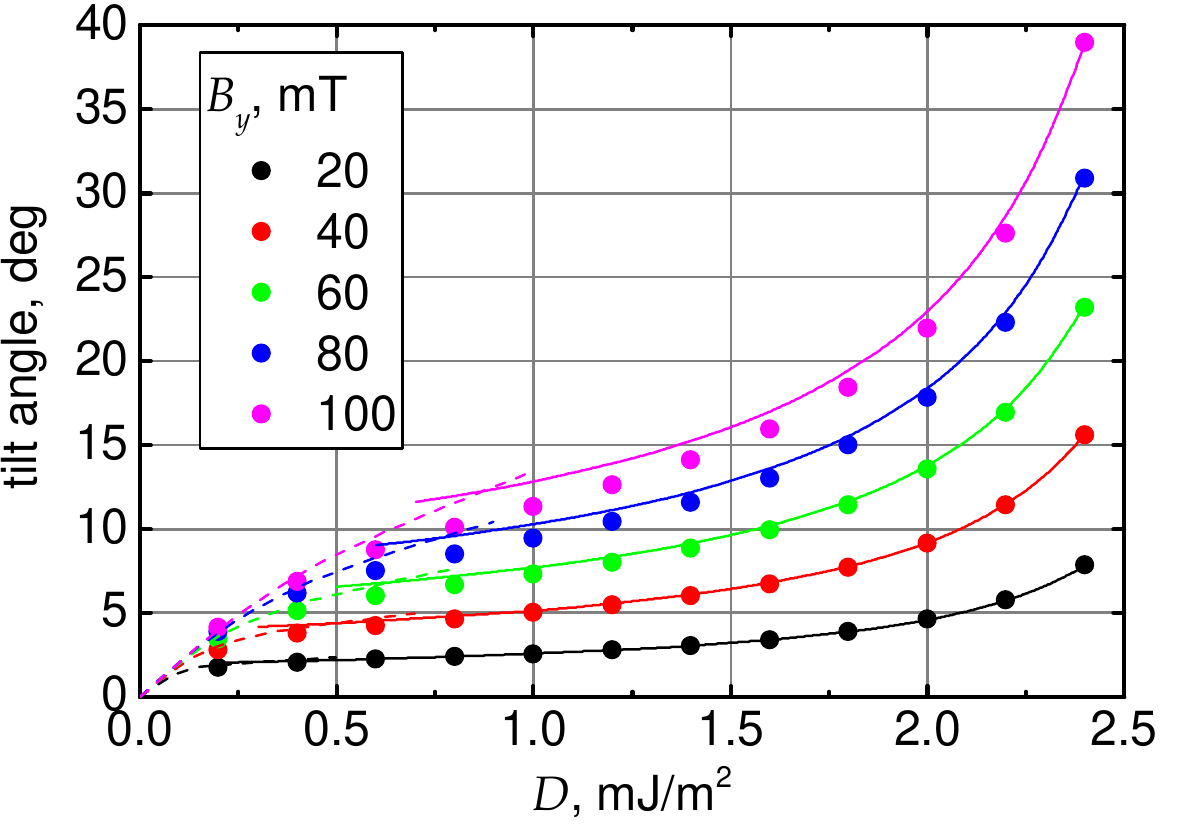}
  \caption{Equilibrium tilt angle, $\beta$, vs. DMI strength $D$ for several
    values of the applied field $B_y = \mu_0 H$, alongside with the
    predictions of Eq.~\eqref{titltalin} (solid lines) and
    Eq.~\eqref{betabis} (dashed lines).}
  \label{fig:tiltvsdmi}
\end{figure}

For lower fields $h$, the agreement with the predictions of the
analytical theory becomes much better. We illustrate this by
presenting the results of the full two-dimensional numerical
simulations against the analytical predictions by
Eqs.~\eqref{titltalin} and \eqref{betabis} for smaller fields in the
whole range of values of $\kappa$. Figure~\ref{fig:tiltvsfield} shows
the dependence of the equilibrium tilt angle $\beta$ on the
applied field for several values of the DMI strength. As can be seen
from the figure, the agreement between the theory and the
numerics rapidly increases as the applied magnetic field or the DMI
strength are decreased. This trend can also be seen from the plot of
the equilibrium tilt angle as a function of the DMI strength for
several value of the applied field shown in Fig.~\ref{fig:tiltvsdmi}.

\begin{figure}[t]
  \centering
  \includegraphics[width=3.15in]{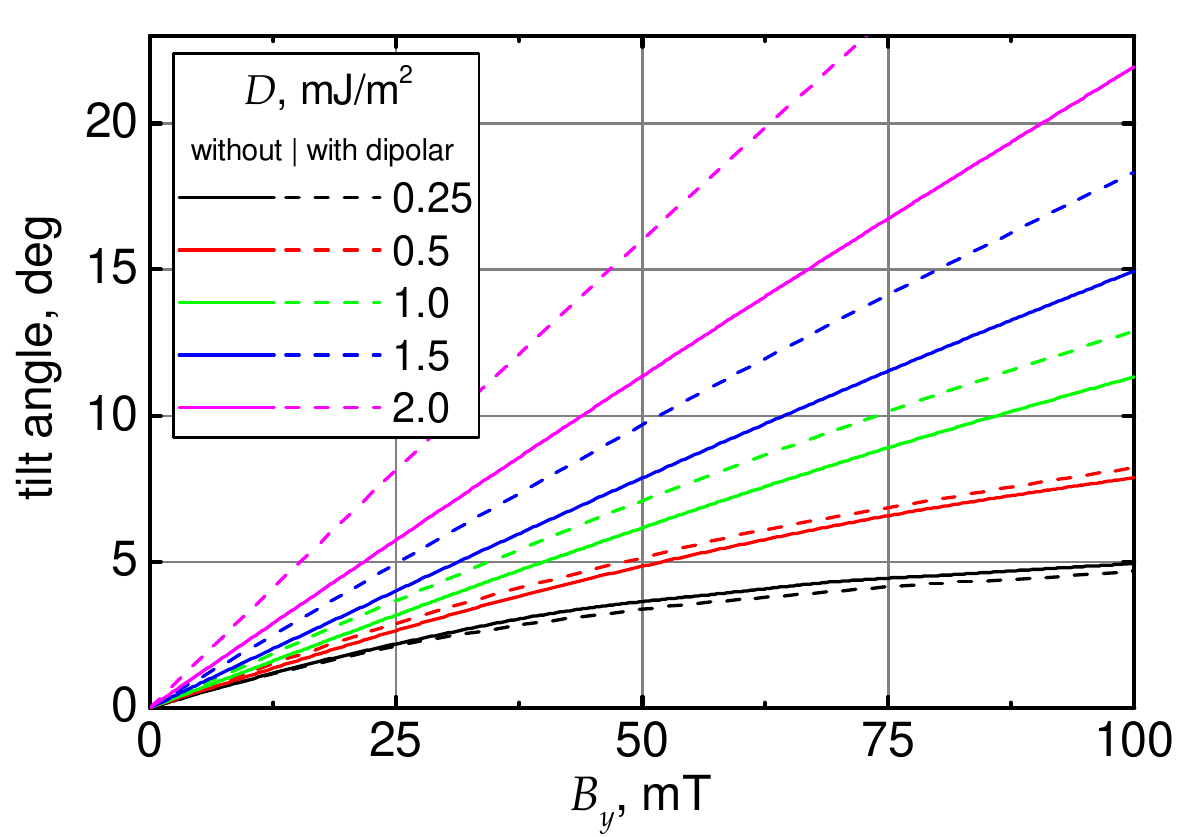}
  \caption{Equilibrium tilt angle, $\beta$, vs. magnetic field $B_y$
    from simulations with and without the dipolar interactions.}
  \label{fig:tilt_dipolar_By}
\end{figure}

So far we presented the results of micromagnetic simulations for a
nanostrip, using a common approximation that neglects the dipolar
interaction. \cite{Winter1961,Gioia97} More precisely, in the preceding
simulations, the effect of dipolar interactions was accounted only by
introducing a local shape anisotropy term. Let us conclude this
section by discussing the results of micromagnetic simulations for
thin nanostrips with the full account of magnetostatic
interaction. The material parameters and the corresponding
dimensionless parameters were the same as in the simulations without
the dipolar interactions. The length of the nanostrip was extended
$1.5$ times to reduce the effect of magnetic charges at the ends of
the nanostrip. The full accounting of the magnetostatic energy in the
simulations leads to an increase in the DW tilt angle, see
Fig.~\ref{fig:tilt_dipolar_By}. One can see that for the considered
parameters the presence of the dipolar interaction affects the tilt
angle relatively weakly for moderate applied fields and DMI
strengths. Notably, the impact of the dipolar interaction is
practically negligible for sufficiently small DMI strengths, but
increases with increasing DMI strength. We attribute this phenomenon
to a decrease in the DW stiffness as the value of $\kappa$ is
increased, making the domain wall more susceptible to the presence of
the dipolar interactions.

\section{Conclusions}
\label{sec:conclusions}

We have developed an analytical theory of the
Dzyaloshinskii domain wall tilt in a ferromagnetic nanostrip in the
perpendicular in-plane magnetic fields. This type of DW tilt is a
vivid manifestation of the presence of interfacial DMI in ultrathin
ferromagnet/heavy-metal layered structures. Our theory focuses on the
geometric aspect of the problem and treats the DW as a curve, whose
equilibrium shape is determined by minimizing an appropriate geometric
energy functional.

The main ingredients in our theory are the energy densities of edge
and interior domain walls. The former are computed explicitly, and the
latter can be obtained for any given set of parameters, using a
straightforward numerical procedure. We have explicitly considered two
regimes: the regime when the dimensionless magnetic field $h$ is much
smaller than the dimensionless DMI strength $\kappa$ and the regime
when they are both small and comparable. In both regimes, we have
found very good agreement with the micromagnetic simulations for the
tilt angle. 

Our theory has three main findings: First, we derived an exact 1D
domain wall profile for any strength of perpendicular in-plane
magnetic field. Second, we proved that the DW is always a tilted
straight line. Third, this allowed us to obtain an explicit expression
for the DW tilt angle. Moreover, in the wide range of DMI strength (as
long as DW does not develop yet helicoidal structure), we find that
the DW configurations are in general neither N{\'e}el nor Bloch type,
and that the DW energy is anisotropic (depends on the tilt angle).

In the regime of small fields $h \ll \kappa \lesssim 1$, we have found
that the equilibrium angle is proportional to the field strength
[Eq.~\eqref{titltalin}]. On the other hand, for small DMI strengths
the tilt angle exhibits a strongly nonlinear dependence on the field
strength, even for relatively small fields
[Eq.~\eqref{betabis}]. Surprisingly, we found that when
$h \ll \kappa \ll 1$ the equilibrium tilt angle becomes independent of
the DMI strength [Eq.~(\ref{betabis1})], which can be a good
experimental test for our theory.  Equally surprisingly, in the
opposite regime $\kappa \ll h \ll 1$, we have shown that the
equilibrium tilt angle becomes independent of the applied magnetic
field [Eq.~(\ref{betabis2})].

Our results indicate that for moderate DMI strengths the tilt angle
may be used to directly assess the value of the interfacial DMI
constant experimentally.  In other words, our theory gives a method to
infer the DMI constant from the tilt angle measurements of the
Dzyaloshinskii domain wall.  We, therefore, propose an experimental
method that requires only a technique for observing the magnetic
structure under external field (e.g. Kerr microscopy). To improve the
accuracy of the DMI determination, one should measure the tilt angle
as a function of magnetic field $B_y$, as shown in
Fig.~\ref{fig:tiltvsfield}, and fit this experimental curve to our
theory [Eqs.~\eqref{titltalin} or~\eqref{betabis}, depending on
smallness of the DMI strength relative to the magnetic field].

\begin{acknowledgments} 

  We thank O. Tchernyshyov for helpful discussions. C.\,B.\,M. was
  supported, in part, by NSF via grant DMS-1614948. V.\,S. would like
  to acknowledge support from EPSRC grant EP/K02390X/1 and Leverhulme
  grant RPG-2014-226. O.\,A.\,T. acknowledges support by the
  Grants-in-Aid for Scientific Research (No.\,25247056, No.\,15H01009,
  No.\,17K05511, and No.\,17H05173) from the Ministry of Education,
  Culture, Sports, Science and Technology (MEXT) of Japan; JSPS-RFBR
  grant; and MaHoJeRo grant. V.S. is grateful to Basque Center for
  Applied Mathematics (BCAM) for its hospitality and support.

\end{acknowledgments}

\bibliography{tilt_DW_references}

\end{document}